\documentclass[structabstract]{aa}  
\usepackage{graphicx}
\usepackage{txfonts}
\begin{document}

   \title{Discovery of diffuse radio emission in the galaxy cluster A1689}

   \subtitle{}

 \author{ V. Vacca \inst{1,2},
          F. Govoni \inst{2},
          M. Murgia \inst{2},
          G. Giovannini \inst{3,4},
          L. Feretti \inst{4},
          M. Tugnoli  \inst{4},
          M.A. Verheijen \inst{5}, \and 
          G.B. Taylor \inst{6,7}  
          }

  \institute{Dipartimento di Fisica, Universit\`a degli Studi di Cagliari, 
    Cittadella Universitaria, I-09042 Monserrato (CA), Italy
         \and
         INAF - Osservatorio Astronomico di Cagliari,
             Strada 54, Loc. Poggio dei Pini, 09012 Capoterra (Ca), Italy
          \and
             Dipartimento di Astronomia, Universit\`a degli Studi di Bologna, 
             Via Ranzani 1, 40127 Bologna, Italy
         \and             
             INAF - Istituto di Radioastronomia, Via P.Gobetti 101, 
             40129 Bologna, Italy 
         \and
             Kapteyn Astronomical Institute, University of Groningen, 
             Landleven 12, Groningen, 
             The Netherlands 
         \and    
            Department of Physics and Astronomy, University of New Mexico, MSC 07 4220, Albuquerque, New Mexico, USA
         \and
              Adjunct Astronomer at the National Radio Astronomy Observatory, 
              Socorro, NM 87801 USA.
                }

  \date{Received MM DD, YY; accepted MM DD, YY}

% \abstract{}{}{}{}{} 
% 5 {} token are mandatory
 %
  \abstract
  {}
  {The aim of this work is to investigate the possible presence of 
  extended diffuse synchrotron radio emission associated with the 
  intracluster medium of the complex galaxy cluster A1689.}
  {The radio continuum emission of A1689 has been investigated
  by analyzing archival observations at 1.2 and 1.4\,{\rm GHz} obtained with the
  Very Large Array in different configurations.}
  {We report the detection of an extended, diffuse, low-surface brightness 
  radio emission located in the central region of A1689. The surface brightness profile
  of the diffuse emission at 1.2\,{\rm GHz} indicates a central radio brightness of 
  $\simeq$1.7\,$\mu${\rm Jy/arcsec}$^2$ and the 3$\sigma$ radio isophothes reveal the largest 
  linear size to be 730\,kpc.
  Given its central location, the low-level surface brightness, and the comparatively large extension, 
  we classify the diffuse cluster-wide emission in A1689 as a small radio halo. 
  }
  {}


   \titlerunning{Discovery of diffuse emission in the galaxy cluster A1689}
   \authorrunning{V. Vacca et al.}
   \maketitle

\begin{table*}
\caption{Details of the VLA observations of A1689.}             
\label{table1}      
\centering          
\begin{tabular}{cccccccc}     % 8 columns 
\hline\hline       
 R.A.       & Decl.        &  Obs. frequency, $\nu$     & Bandwidth    &     VLA Configuration  &   Time   &   Date  & Project\\
(J2000)     & (J2000)      &    ({\rm MHz})             &  ({\rm MHz}) &        ~               &    (h)   &     ~   &   ~    \\
\hline
13 11 30.2  &  $-$01 20 10.6 & 1365/1435 & 25.0     &      A                  & 3.0     &  Mar 29,30,31 2002 &  AI0098\\
13 11 30.2  &  $-$01 20 35.4 & 1193     & 12.5     &      C                   & 6.0      &  Jan 13, 1999    &  AZ0111\\
13 11 30.2  &  $-$01 20 35.4 & 1193     & 12.5     &      DnC                & 7.8      &  Feb 03,05 1999    &  AZ0111\\
\hline 
\multicolumn{8}{l}{\scriptsize Col. 1, Col. 2: Pointing position; Col. 3: Observing frequency; Col. 4: Observing bandwidth; Col. 5: VLA configuration; Col. 6: Time on source;}\\
\multicolumn{8}{l}{\scriptsize Col. 7: Dates of observations; Col. 8: VLA project name.}\\     
\end{tabular}
\end{table*}

\section{Introduction}
According to the hierarchical scenario of structure formation, 
massive galaxy clusters form from the merger of 
galaxy groups and subclusters. 
During these mergers, gravitational energies are released
that are high as $\gtrsim 10^{64}$\,{\rm ergs} (e.g., Sarazin 2002). 
This significant amount of energy drives shocks and turbulence 
into the thermal intracluster medium (ICM): the energy is injected on large 
spatial scales and then cascades turbulently to smaller scales. 
Shocks and the turbulence associated with a major cluster merger event are 
thought to accelerate particles and compress magnetic field in the ICM 
(e.g. Roettiger et al. 1999; Skillman et al. 2008; Skillman et al. 2011). These nonthermal components are 
expected to lead to large-scale diffuse synchrotron emission associated 
with the ICM -- an expectation confirmed by sensitive radio observations. 
In fact, in an increasing number of merging galaxy clusters,
 diffuse sources known as radio halos and radio relics, which have no obvious optical counterpart, 
have been detected (Feretti \& Giovannini 2008, Ferrari et al. 2008). 

Radio halos and radio relics are diffuse, low-surface-brightness ($\sim$ 1\,$\mu${\rm Jy
arcsec}$^{-2}$ at 1.4\,{\rm GHz}), steep-spectrum\footnote{We use the convention $S_{\nu}\propto \nu^{-\alpha}$.} 
($\alpha\gtrsim 1$) synchrotron sources with linear size $\gtrsim$\,Mpc. The radio halo emission permeates the central 
volume of some galaxy clusters, is generally characterized by a regular morphology, and is not polarized. 
 Instead, radio relic emission is located in the galaxy cluster
outskirts, usually filamentary, and polarized (10--30\%).
While radio halos are thought to be sustained mainly by turbulence due to cluster mergers 
(e.g, Brunetti \& Lazarian 2007), radio relics are supposed to trace the shock waves 
generated during these processes of cluster formation (e.g., Ensslin et al. 1998).

Radio halos offer a unique chance to study the nonthermal components of the ICM in the 
central region of the clusters on Mpc scales but, unfortunately, their faintness means their statistics remain poor.
To date, only about 30 radio halos are known (Giovannini et al. 2009).
They are all found in clusters displaying features of intense merger
activity (Buote 2001, Feretti 2002, Govoni et al. 2004, Cassano et al. 2010, Rossetti et al. 2011), but not all merging
clusters appear to host a radio halo.  X-ray and radio properties of
the clusters hosting such sources seem to be intimately connected, as
indicated by the similarity of the X-ray and radio morphology, 
 by the 
correlation between the radio halo power and X-ray
cluster temperature (Liang et al. 1999, Colafrancesco 1999)
and by
the strong correlation between the radio halo power $P_{\rm
  1.4\,GHz}$ at 1.4\,{\rm GHz} and the X-ray cluster luminosity
$L_{\rm X}$ (Feretti 2002, Brunetti et al. 2007, Giovannini et al. 2009).  
Actually, recent
deep radio observations also revealed the presence of radio halos in
clusters characterized by low X-ray luminosity (e.g. 0217+70: Brown
et al. 2011; A523: Giovannini et al. 2011), a  more
complex situation than described by the present models and numerical simulations
of radio halo formation.

While radio halos are typically found in merging clusters, few
relaxed, cool-core, galaxy clusters exhibit signs of diffuse
synchrotron emission that extends far from the dominant radio galaxy at 
the cluster center, forming a so-called mini-halo.
Burns et al. (1992) was the first to recognize a distinction between cluster-wide
halos, such as in Coma, and mini-halos associated with cool cores, such as in Perseus.
Mini-halos are extended on a moderate scale (typically $\simeq$ 500\,{\rm kpc}) and,
in common with the large-scale halos observed in merging clusters of
galaxies, have a steep spectrum and very low surface
brightness. Their relatively small angular size in combination with,
in many cases, strong emission of the central radio galaxy,
complicates their detection, thus our current observational knowledge
on mini-halos is limited to only a handful of clusters.  The origin
and physical properties of mini halos are still poorly known
(Gitti et al. 2004, Govoni et al. 2009).  We note that, although
cooling core clusters are generally considered relaxed systems, when
analyzed in detail they sometimes reveal peculiar X-ray features in
the cluster center, which may indicate a link between the
mini-halo emission and some minor merger activity.  Indeed, Burns et
al. (2008) have simulated the formation of both cool core and non-cool core
clusters in the same numerical volume.  These simulations confirm
that non-cool clusters are formed via major mergers early in their
history, which destroyed the cool cores and left significant residual
kinetic energy in the gas that might be used to power the radio
halos.  In contrast, cool core clusters do not suffer any major
mergers, thus preserving the central cool regions.  However, the cool
core clusters do experience regular smaller mergers, which still inject
energy into the intracluster medium, but more modestly than in the
non-cool core clusters.  These minor mergers might power the
mini-halos.

Further studies are required to better understand extended diffuse
synchrotron radio sources and their connection with the dynamical
state of the cluster. In this context, as a part of an ongoing program
aimed at finding new diffuse sources in galaxy clusters, we
investigate the radio emission of the galaxy cluster A1689.

In this work we present the detection of a diffuse synchrotron radio
source in this complex galaxy cluster.  In \S\,\ref{A1689 properties} we 
summarize the cluster properties. In \S\,\ref{Radio observations
  and data reduction} we describe the radio observation and the data
reduction.  In \S\,\ref{A new diffuse emission in A1689} we show the
results and we attempt to classify this diffuse radio
emission.  Finally, in \S\,\ref{Conclusion} we draw some conclusions.

Throughout this paper we adopt a $\Lambda$CDM cosmology with $H_{\rm
  0}=71$\,{\rm km s}$^{\rm -1}$\,{\rm Mpc}$^{\rm-1}$, $\Omega_{\rm
  m}=0.27$, and $\Omega_{\rm \Lambda}=0.73$. At the distance of A1689
(z=0.1832; Struble \& Rood 1999), the luminosity distance is $D_{\rm
  L}$=880\,{\rm Mpc}, and 1\arcsec\, corresponds to 3.05\,{\rm kpc}.

\section{A1689 properties}
\label{A1689 properties}
A1689 is a galaxy cluster characterized by a complex optical and 
X-ray structure. 
At first inspection, it has a regular spherical shape and a
strong peak in the surface brightness profile in X-rays (Peres et al. 1998),
features that apparently indicate a cooling core. Indeed,  
Chen et al. (2007) found a short central cooling time for A1689
$t_{\rm cool}=4.7\pm0.4$\,Gyr and a high accretion rate  
${\dot M}=683_{-182}^{239}M_{\odot}\,{\rm yr}^{-1}$.
Its Einstein radius is
the largest known so far (Tyson et al. 1990, Miralda-Escud\'e \& Babul
1995, Clowe \& Schneider 2001, Broadhurst et al. 2005a, Broadhurst et
al. 2005b, Umetsu \& Broadhurst 2008), a characteristic that has made
it attractive for lensing analysis.  
Strong and weak lensing studies
have revealed discrepant mass values.  Differences in the mass
estimations have also been found between lensing and X-ray
observations.  

By means of \emph{XMM-Newton} observations, Andersson
\& Madejski (2004) found evidence of an ongoing merger. 
They found a nonuniform radial temperature 
of the X-ray emitting gas across the cluster, and these features 
suggest a complex dynamical state.
A major merger in which a subcluster moves along the line-of-sight could also 
explain the different mass estimates from X-ray and lensing analysis.  
An alternative explanation of this difference has been given by Lemze et
al. (2008). By using a model-independent approach based on a
simultaneous fit of lensing and X-ray surface brightness data, they
find good agreement between the lensing mass profile and the X-ray
emission profile.  Their model, however, implies a temperature
discrepancy with respect to the one derived solely from X-ray
observations. The authors state that their result could support the
presence of colder, denser, and more luminous small-scale structures
that could be responsible for a bias in the observed temperature
(Kawahara et al. 2007).  Peng et al. (2009) refute this view and
show that X-ray and lensing estimates can be made consistent by
considering a prolate distribution for the gas temperature and density
with the major axis aligned with the line-of-sight. 

By using the Hubble's Advanced Camera for Surveys (ACS) 
and combining the strong and weak lensing approaches, 
Limousin et al. (2007) apply an accurate mass model to described the 
observed substructures, obtaining good agreement
between the strong and weak lensing analysis. To reduce the discrepancy 
between X-ray and lensing approaches, Riemer-S{\o}renson et al. (2009) 
combined this lensing analysis with high-resolution X-ray \emph{Chandra} 
temperature and hardness ratio images. They find consistent mass profiles 
and indications that the main clump is in a hydrostatical equilibrium.
By means of the same ACS observations,
Coe et al. (2010) trace the most
detailed mass distribution with a resolution of 25\,{\rm kpc}. They
estimate a mass 
$M_{200}=1.8^{\rm +0.4}_{\rm -0.3}\times10^{\rm 15}M_{\odot}\,h^{\rm -1}_{\rm 70}$ 
within a radius 
$r_{200}=2.4^{\rm +0.1}_{\rm -0.2}{\rm Mpc}\,h^{\rm -1}_{\rm 70}$. In addition, their
results agree with the presence of a high mass along
the line-of-sight, confirming previous studies present in the literature.
Analyzing the same ACS dataset, Leonard et al. (2011) detect two
dominant peaks of mass concentration that they identify as possible
remnants of a recent merger.

The head-on merger scenario also seems supported by optical observations.
In fact, the central region of the cluster lacks 
a dominant galaxy at its centroid, while it hosts multiple nuclei/galaxies 
within a single optical halo and possibly merging. Indeed,
Girardi et al. (1997) find that A1689 
appears to consist of three distinct groups of galaxies possibly aligned and 
separated well in velocity, suggesting a merger along the line of sight. 
Such a geometry of the merger leads to an apparent enhancement of the observed 
velocity dispersion that, by using the temperature measured by 
White \& Fabian (1995), they find to be 1350\,km~s$^{-1}$.

\begin{figure*}
   \centering
  \includegraphics[width=13cm]{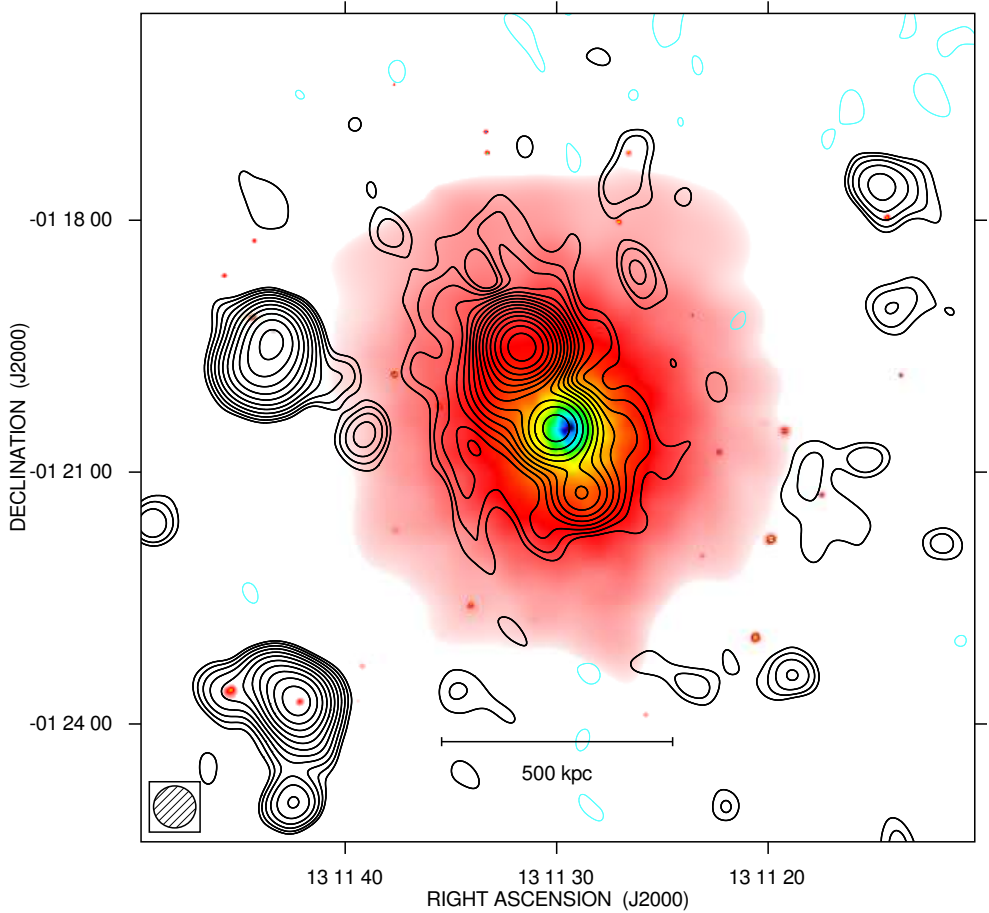}
      \caption{Total intensity radio contours at 1.2\,{\rm GHz}  (VLA in DnC configuration) with an FWHM of $30\arcsec \times 30\arcsec$.  The contour levels are drawn at -3$\times10^{\rm -4}$\,{\rm Jy/beam}, 3$\times10^{\rm -4}$\,{\rm Jy/beam}, 
and the rest are spaced by a factor $\sqrt{2}$. The sensitivity (1$\sigma$) is 1$\times10^{\rm -4}$\,{\rm Jy/beam}. The contours of the radio intensity are overlaid on a \emph{Chandra} X-ray (Obs. id: 6930) 0.1$-$10\,{\rm keV} band, adaptively smoothed count image. }
              \label{radioX}
    \end{figure*}

 \begin{figure*}
   \centering
  \includegraphics[width=13cm]{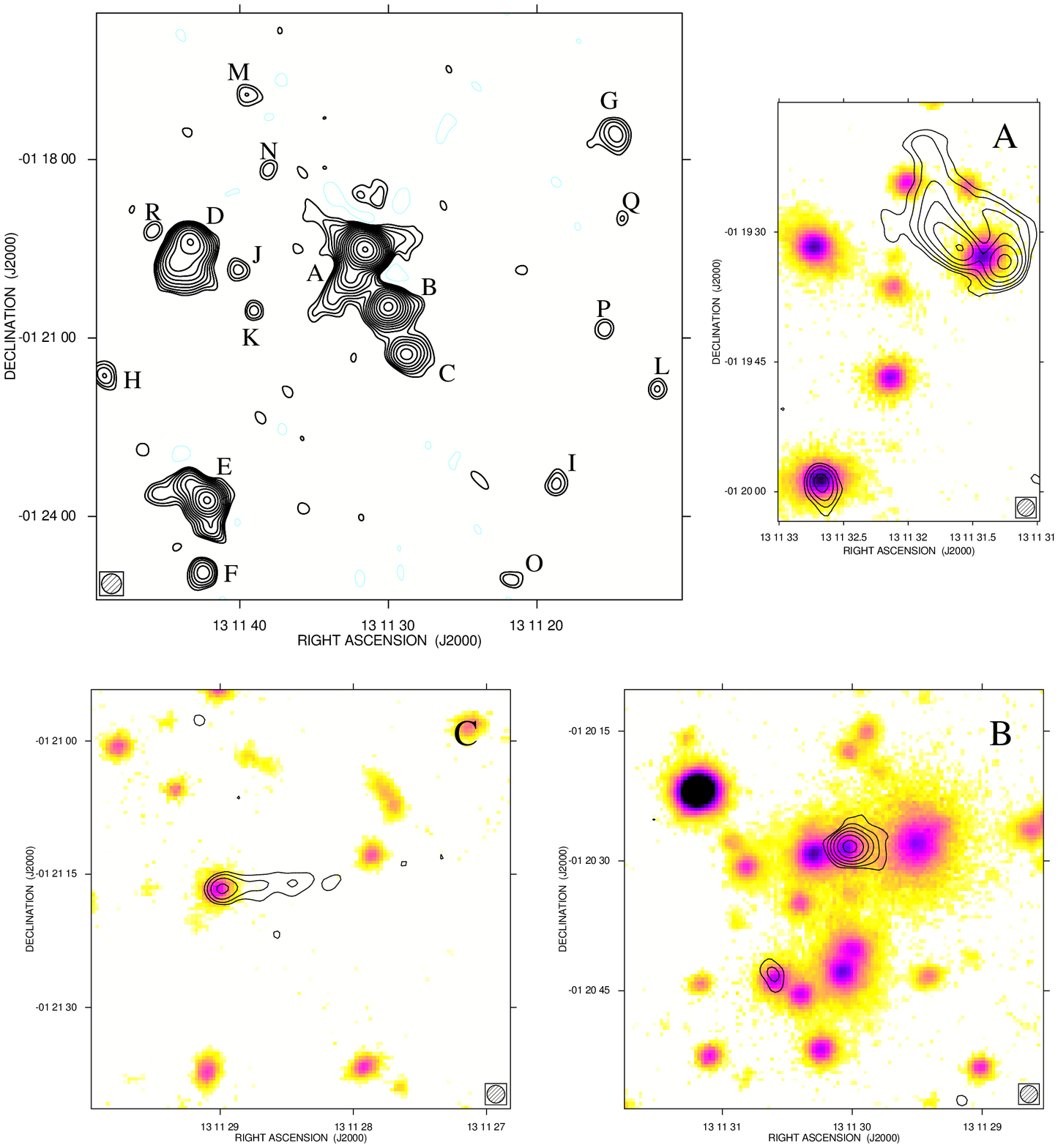}
      \caption{\emph{Top left panel}: Total intensity radio contours at 1.2\,{\rm GHz}  (VLA in C configuration) with an FWHM of $20\arcsec \times 20\arcsec$.  The contour levels are drawn at -3.3$\times10^{\rm -4}$\,{\rm Jy/beam}, 3.3$\times10^{\rm -4}$\,{\rm Jy/beam}, 
and the rest are spaced by a factor $\sqrt{2}$. The sensitivity (1$\sigma$) is 1.1$\times10^{\rm -4}$\,{\rm Jy/beam}. \emph{Bottom left, right panels}: Total intensity radio contours at 1.4\,{\rm GHz}  (VLA in A configuration) with an FWHM of $2\arcsec \times 2\arcsec$.  The first contour level is drawn at 13.5$\times10^{\rm -5}$\,{\rm Jy/beam}, and the rest are spaced by a factor 2. The sensitivity (1$\sigma$) is 4.5$\times10^{\rm -5}$\,{\rm Jy/beam}. The contours of the radio intensity are overlaid on the red plate of the Sloan Digital Sky Survey.}
              \label{radioott}
    \end{figure*}

\section{Radio observations and data reduction}
\label{Radio observations and data reduction}
We present archival observations of A1689 at 1.2 and 1.4\,{\rm GHz} performed
at the Very Large Array (VLA) in spectral line mode in the A, C, and DnC
configurations.  The details of the observations are summarized in
Table\,\ref{table1}.  The data were reduced following standard
procedures using the NRAO's Astronomical Image Processing System
(AIPS) package.  Surface brightness images were produced using 
the AIPS task IMAGR.  In the C and DnC configuration the data were
collected with a total bandwidth of 12.5\,{\rm MHz}, subdivided into seven
channels with a band-with of $\sim$1.6\,{\rm MHz} each.  The source
1331+305 (3C286) was used as flux and bandpass calibrator.  
The nearby source 1354$-$021 was observed for complex gain calibration. 
Radio interference was carefully excised channel by channel, and several
cycles of self-calibration and imaging were applied to remove residual
phase variations.  We averaged the seven channels together in the
gridding process using IMAGR.  The A configuration was calibrated by
using the source 1331+305 as flux calibrator, and the nearby source
1246-075 as phase reference.  The two IFs were averaged to obtain the
surface brightness image.

\section{Diffuse emission in A1689}
\label{A new diffuse emission in A1689}

The radio iso-contours at 1.2\,{\rm GHz} of A1689 are shown in
Fig.\,\ref{radioX}.  This image was obtained with the VLA in DnC
configuration, and it was convolved with a circular FWHM beam of
30\arcsec.  To compare the radio and X-ray cluster emission, the radio
contours are overlaid on the \emph{Chandra} image in the 0.1$-$10\, {\rm
  keV} band.  We find that the central region of A1689 is permeated by
a low-surface brightness diffuse emission with a few discrete sources
embedded.  The radio morphology of the diffuse emission is quite
regular with a round shape, following the regular structure of the
cluster X-ray emission.  As measured from the 3$\sigma$ radio
isophote, the overall diffuse emission has an angular extension of
about 4\arcmin~(the Largest Linear Size is LLS$\simeq 730$\,{\rm kpc} at the cluster distance).
However, as pointed out in Murgia et al. (2009), we note that the size
of the diffuse emission calculated from the contour levels should be
considered carefully, since it depends on the sensitivity of the
radio image.
\begin{figure}
   \centering
  \includegraphics[width=9cm]{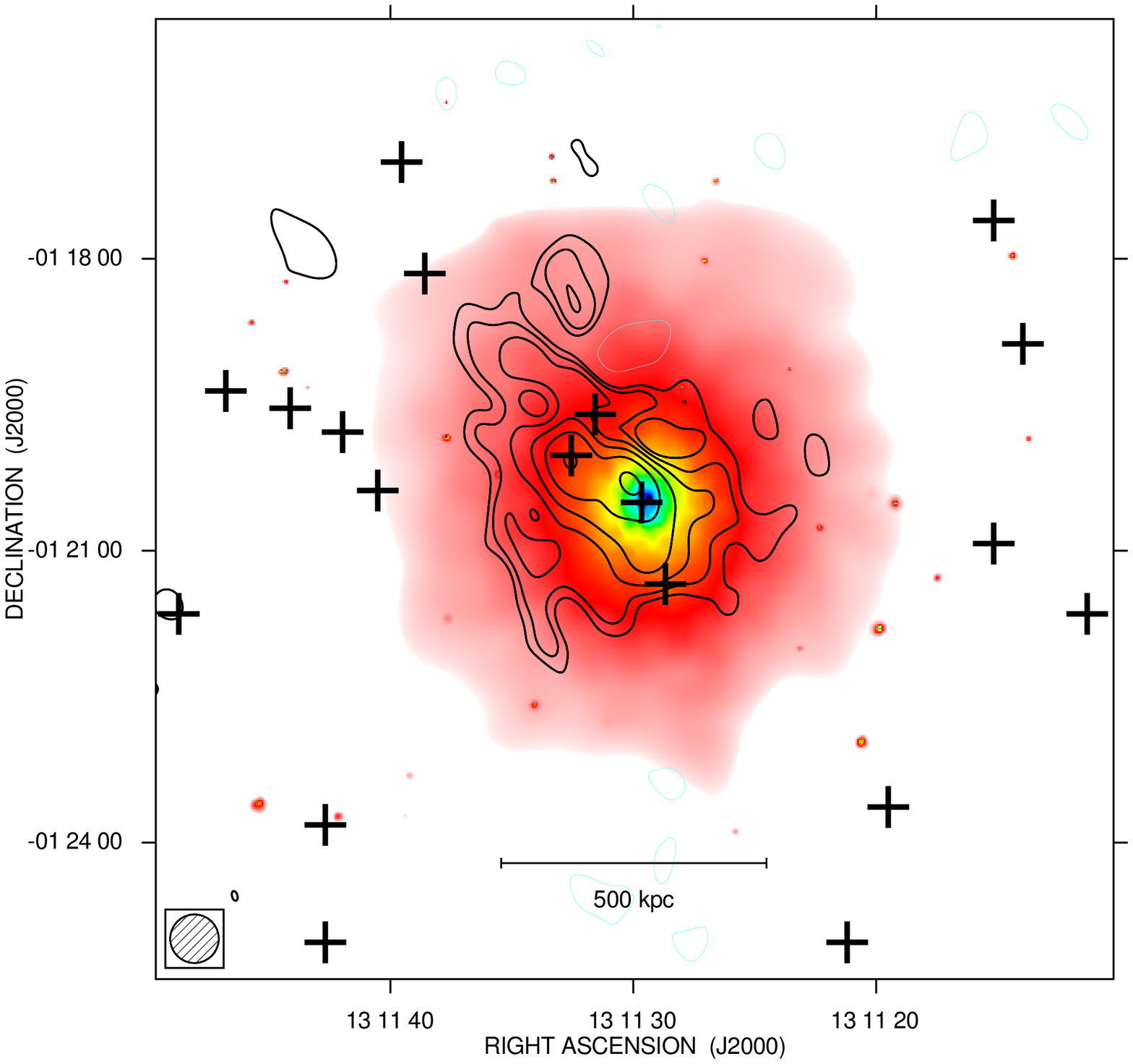}
      \caption{Total intensity radio contours at 1.2\,{\rm GHz}  with the VLA in DnC configuration after subtraction of discrete sources. The image has an
FWHM of $30\arcsec \times 30\arcsec$. The contour levels are drawn at -3$\times10^{\rm -4}$\,{\rm Jy/beam}, 3$\times10^{\rm -4}$\,{\rm Jy/beam}, 
and the rest are spaced by a factor $\sqrt{2}$. The sensitivity (1$\sigma$) is 1$\times10^{\rm -4}$\,{\rm Jy/beam}. Crosses indicate the positions of the subtracted discrete sources. The contours of the radio intensity are overlaid on the \emph{Chandra} X-ray image shown in Fig.\,\ref{radioX}. 
}
              \label{subtraction}
    \end{figure}
\begin{figure}
   \centering
  \includegraphics[width=9cm]{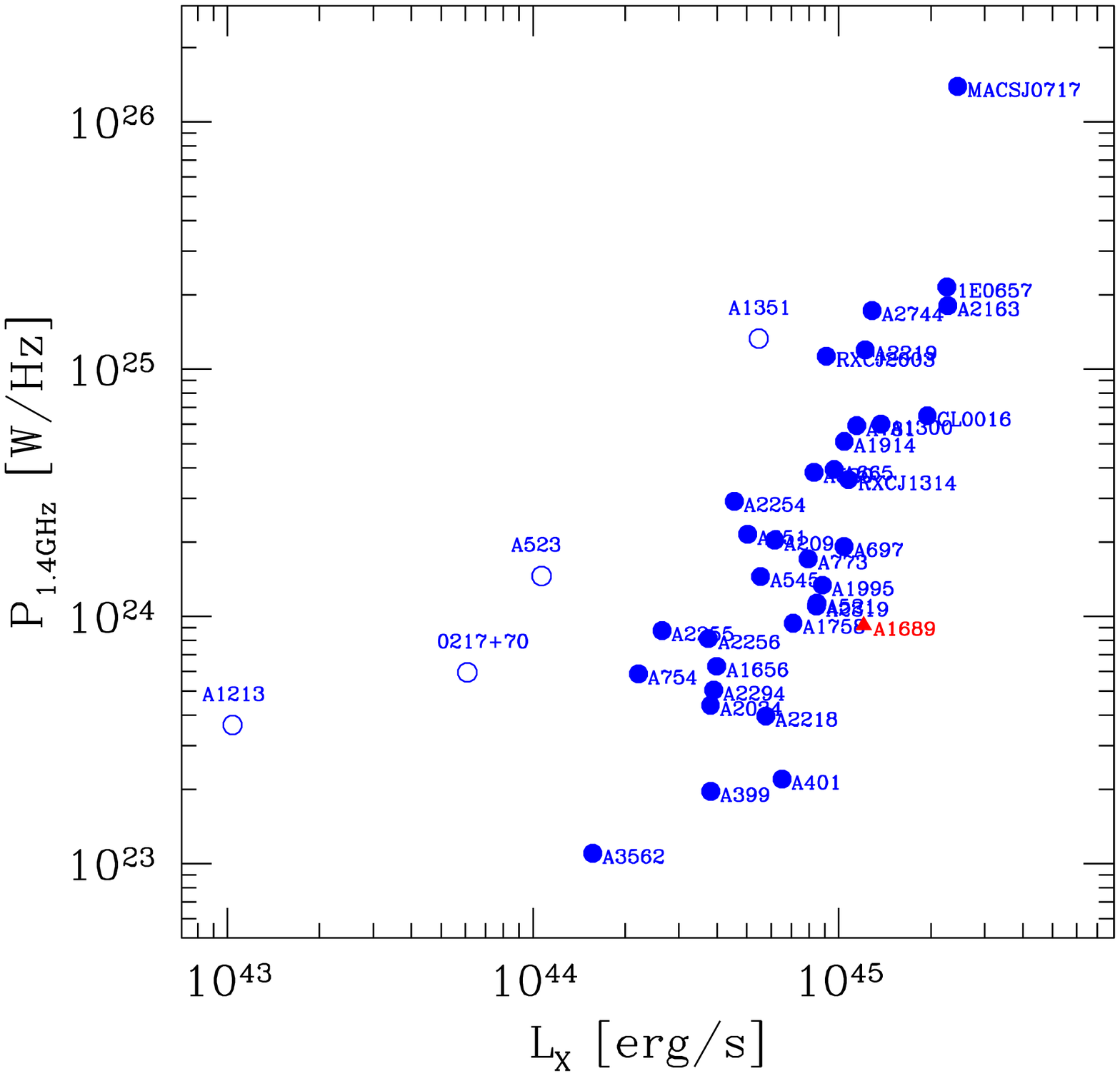}
      \caption{Relation between the halo radio power at 1.4\,{\rm GHz} and
the cluster X-ray luminosity between 0.1 and 2.4\,{\rm keV}. A1689 is indicated with a red triangle. 
Blue dots are classical radio halos, while empty blue dots are outlier clusters. The data are taken from 
Giovannini et al. (2009) and references therein, with the addition of 
CL0016+16 (Giovannini \& Feretti 2000), MCSJ0717+3745 (Bonafede et al. 2009), 
A399 (Murgia et al. 2010), A781 (Govoni et al. 2011),
A523 (Giovannini et al. 2011), and 0217+70 (Brown et al. 2011).}
              \label{relation}
    \end{figure}

To separate the diffuse radio emission from discrete sources, we
produced images at higher resolution. In Fig.\,\ref{radioott} we
present the radio contours of A1689 at 1.2\,{\rm GHz} taken with the VLA in C
configuration, convolved with an FWHM beam of 20\arcsec.  The discrete
sources are labeled in the figure.  A few discrete radio sources
(labeled A, B, C) are embedded in the diffuse cluster emission, and
their positions and flux densities are given in Table\,\ref{flux}.
Source B is located close to the cluster X-ray peak.  In the
bottom and righthand panels of Fig.\,\ref{radioott}, their radio 
contours at 1.4\,{\rm GHz} taken with the VLA in A configuration (convolved with an
FWHM beam of 2\arcsec) are overlaid to the optical Sloan Digital Sky
Survey.  From this figure it is evident that source A, when observed at
high resolution, is indeed composed of two distinct discrete sources: 
a wide-angle tailed radio source and a point source, 
shown respectively in the top right and bottom lefthand corners of 
the top righthand panel of Fig.\,\ref{radioott}
(the flux density of source A given in Table\,\ref{flux} includes both
sources).  All these discrete sources have an optical counterpart.

To ensure that the large-scale diffuse emission is not caused by the
blending of discrete sources, in Fig.\,\ref{subtraction} we present
the total intensity radio contours at 1.2\,{\rm GHz} with the VLA in DnC
configuration after subtraction of discrete sources. We produced an
image of the discrete sources by using only the longest baselines of
the DnC configuration dataset and uniform weighting.  The clean
components of this image were then subtracted in the ($u,v$) plane by using
the AIPS task UVSUB.  The image with the discrete sources subtracted
confirms the presence of a low-surface brightness diffuse radio
emission at the cluster center.  The radio contours are overlaid
on the \emph{Chandra} X-ray image presented in Fig.\,\ref{radioX}.  Crosses
indicate the positions of the subtracted discrete sources.

The total flux density is calculated from the DnC dataset after 
a primary beam correction by integrating the total intensity 
surface brightness in the region of the diffuse emission
down to the 3$\sigma$ level.  The resulting total flux density
at 1.2\,{\rm GHz} is estimated to be $\simeq$ (91.6$\pm$2.7)\,{\rm mJy}.
By subtracting the flux density of the embedded discrete 
sources A, B, and C as derived in the
C configuration dataset (see Table\,\ref{flux}),
a flux density of $\simeq$ (11.7$\pm$3.4)\,{\rm  mJy} appears to be associated 
with the low-brightness diffuse emission.
This flux density value corresponds to a radio power of
$P_{\rm 1.2\,GHz} = 1.08\times10^{24}$\,{\rm W~Hz}$^{\rm -1}$. 
To compare this result with studies of other radio halos, 
we rescaled the 
 corresponding radio power at 1.4\,{\rm GHz} $P_{\rm 1.4\,GHz} = 9.21\times10^{23}$\,{\rm W~Hz}$^{\rm -1}$, correcting by a factor 0.85 as estimated assuming a spectral index $\alpha= 1$. 

A1689 is part of an X-ray flux-limited galaxy cluster 
sample selected from the ROSAT All-Sky Survey by 
Reiprich \& B{\"o}hringer (2002). Its X-ray luminosity 
(corrected for our cosmology) in the 0.1-2.4\,{\rm keV} band
is $1.2\times 10^{\rm 45}$\,{\rm erg/sec}. 
Therefore, the radio power $P_{\rm 1.4\,GHz}$, 
the radio's largest linear size ($LLS$),  
and the X-ray luminosity ($L_{\rm X}$) of A1689 agree with the 
$P_{\rm 1.4\,GHz}-LLS$ and $P_{\rm 1.4\,GHz}-L_{\rm X}$ relations 
known for the other halos in clusters (Giovannini et al. 2009).

In Fig.\,\ref{relation} we plot the radio power calculated at
1.4\,{\rm GHz} versus the 0.1$-$2.4\,{\rm keV} X-ray luminosity for
clusters hosting radio halos.  Classical
powerful radio halos in X-ray luminous clusters have been found
to show a correlation between the radio power and the X-ray
luminosity.  A1689 appears consistent with the relation
found for other radio halos. The outliers
are the few known radio halos over-luminous in radio
with respect to the empirical radio - X-ray correlation, revealing a
complex scenario for the radio halo formation (Giovannini et
al. 2011).

\begin{table}
\caption{Information on discrete radio sources embedded 
in the central diffuse cluster emission.}
\begin{center}
\begin{tabular} {lccl} 
\hline
Label   &  R.A.      &  Decl.       &  S$_{\rm 1.2\,GHz}$      \\
               &  (J2000)   &  (J2000)     &  ({\rm mJy})       \\
\hline
A              & 13 11 31.5 & -01 19 31    &  62.0$\pm$ 2.0     \\
B              & 13 11 30.0 & -01 20 28    &  14.0$\pm$ 0.4     \\
C              & 13 11 29.0 & -01 21 17    &   3.9$\pm$ 0.1    \\
\hline
\multicolumn{4}{l}{\scriptsize Col. 1: Source label; Col. 2, 3: Source position (R.A., Decl.);}\\ 
\multicolumn{4}{l}{\scriptsize Col. 4: Source flux density at 1.2\,{\rm GHz}.}\\
\end{tabular}
\label{flux}
\end{center}
\end{table}

\subsection{Azimuthally averaged brightness profile of the radio emission}
\label{Azimuthally averaged brightness profile of the radio emission}
\begin{figure*}
\centering
\includegraphics[width=18 cm]{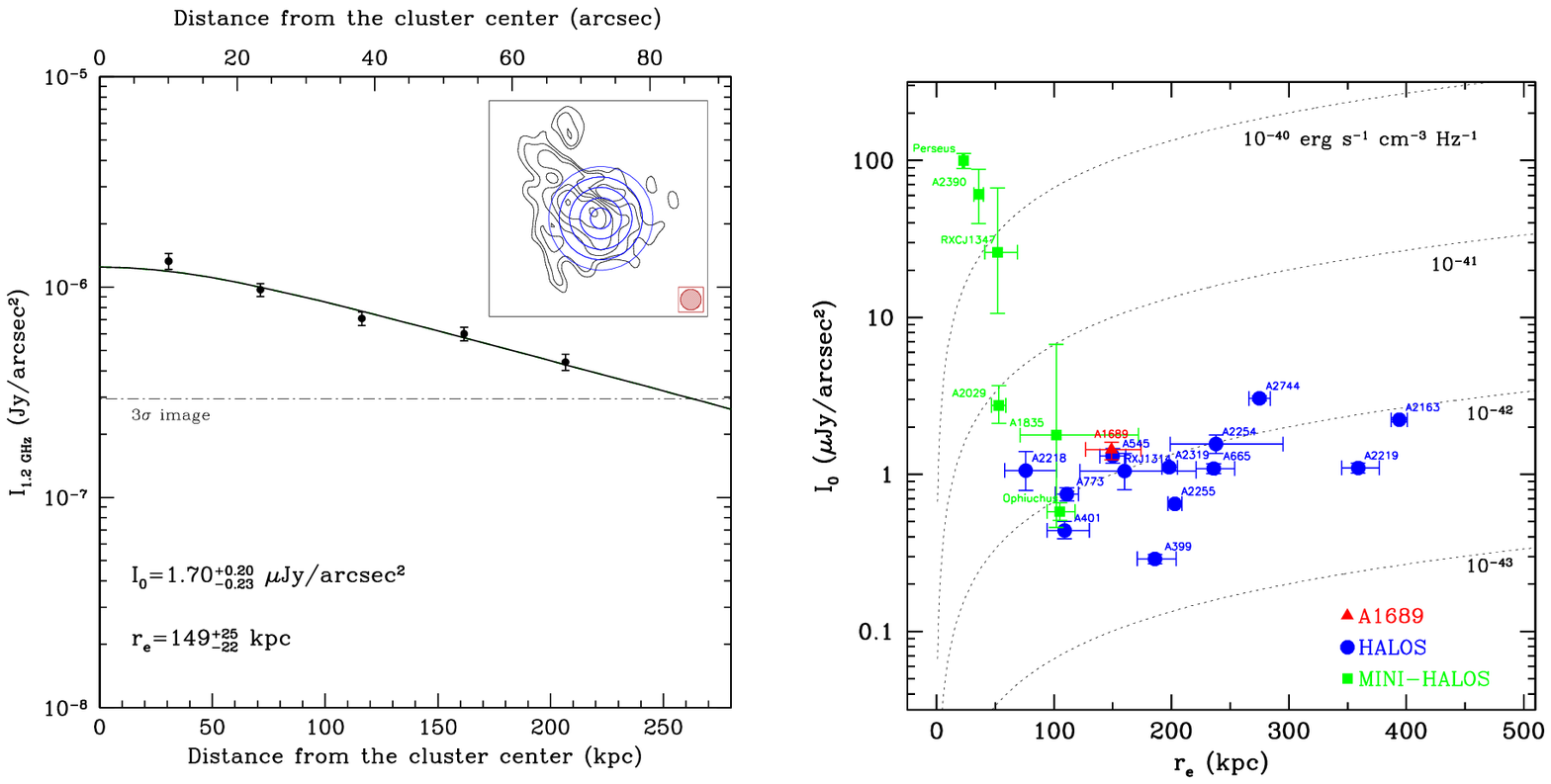}
\caption{
\emph{Left}: The azimuthally averaged brightness profile of the radio halo emission
in A1689 at 1.2\,{\rm GHz}. The profile has been calculated in concentric annuli  centered on the cluster X-ray peak, 
as shown in the inset panel. 
The horizontal dashed-dotted line indicates the 3$\sigma$ noise level of 
the radio image, while the continuous line indicates the best-fit 
profile described by an exponential law (see text). 
\emph{Right}: Best-fit central brightness $I_{\rm 0}$ at 1.4\,{\rm GHz} versus the 
length scale $r_{\rm e}$ of A1689 in comparison with azimuthally 
averaged brightness profiles of radio halos and mini-halos taken from the literature 
(Murgia et al. 2009, Murgia et al. 2010, and references therein).
The dotted 
lines indicate regions of constant synchrotron emissivity. The central brightness of A1689 has been rescaled at 1.4\,{\rm GHz} with $\alpha=1$ (see text).}
\label{profilo}
\end{figure*}

In the lefthand panel of Fig.\ref{profilo} we show the azimuthally averaged 
radio halo brightness profile at 1.2\,{\rm GHz} obtained from the 
DnC configuration image
after subtracting of the discrete sources (see Fig.\ref{subtraction}) 
and correcting for the 
primary beam 
attenuation. Each data point represents the average brightness in 
concentric annuli of half beam width ($\simeq$15\arcsec) 
centered on the cluster X-ray peak. 
The observed brightness profile is 
traced down to a level of 3$\sigma$. 

Following Murgia et al. (2009), we modeled the radio halo brightness
profile, $I(r)$, with an exponential of the form $I(r)=I_{\rm 0}e^{\rm
  -r/r_{\rm e}}$, whose best-fit is shown in the
left panel of Fig.\ref{profilo}.  The proposed method for deriving the
radio brightness, the length scale, hence the radio emissivity of
diffuse sources, is relatively independent of the sensitivity of the
radio observation.  The exponential model is attractive in its
simplicity and involves a minimal set of free parameters. Obviously,
it cannot account for the local deviations from the circular symmetry
of the diffuse emission.

The fit is performed in the image plane as described in Murgia et
al. (2009). To properly take the resolution into account, the
exponential model is first calculated in a 2-dimensional image, with
the same pixel size and field of view as observed, and then
convolved with the same beam by means of a fast Fourier transform. %The
Finally, the model is azimuthally averaged with the same
set of annuli used to obtain the observed radial profile. All these
functions are performed at each step during the fit procedure.  As a
result, the values of the central brightness, $I_{\rm 0}$, and the
e-folding radius $r_{\rm e}$ provided by the fit are deconvolved
quantities, and their estimate includes all the uncertainties related
to the sampling of the radial profile in
annuli of finite width. The fit procedure has been implemented in the
software FARADAY (Murgia et al. 2004).  The best-fit of the
exponential model at 1.2\,{\rm GHz} yields a central brightness \footnote{We checked that the fit of the exponential disk 
performed by masking, and not subtracting, the discrete sources 
is consistent within the uncertainties with the reported values
for $I_{\rm 0}$ and $r_{\rm e}$.} of $I_{\rm
  0}$=($1.70_{-0.23}^{+0.20}$)\,$\mu${\rm Jy/arcsec}$^{\rm 2}$
and $r_{\rm e}$=($149_{-22}^{+25}$)\,{\rm kpc}. The average
radio emissivity over the volume of a sphere of radius 3$r_{\rm e}$
is
\begin{equation}
\langle J \rangle \simeq 7.7 \times 10^{\rm -41} (1+z)^{\rm 3+\alpha} \cdot \frac{I_{\rm 0}}{r_{\rm e}} ~~{\rm(erg\,s^{-1}cm^{-3}Hz^{-1})} 
\label{emissivity}
\end{equation}
where $r_{\rm e}$ and $I_{\rm 0}$ are in units of {\rm kpc} and $\mu${\rm Jy/arcsec}$^{\rm 2}$, respectively.  
From the central brightness and the e-folding radius, we obtained for A1689
an average radio emissivity calculated over the volume of a sphere with radius of 3$r_{\rm e}$,
k-corrected with $\alpha$=1, of
$\langle J \rangle=1.7\times10^{\rm -42}$\,{\rm ergs}$^{\rm -1}${\rm cm}$^{-3}${\rm Hz}$^{-1}$.

In the righthand panel of Fig.\ref{profilo}, we show the best-fit central
brightness $I_{\rm 0}$ rescaled at 1.4\,{\rm GHz} with $\alpha=1$ versus the length scale $r_{\rm e}$ of A1689 in
comparison with the set of radio halos and mini-halos analyzed in
Murgia et al. (2009) and Murgia et al. (2010).  As previously pointed
out, radio halos can have quite different length scales, but their
emissivity is remarkably similar from one halo to another. In
contrast, mini-halos span a wide range of radio emissivity. Some of
them, like Perseus (Pedlar et al. 1990, Burns et al. 1992),
RXJ1347.5-1145 (Gitti et al. 2007) and A2390 (Bacchi et al. 2003) are
characterized by a radio emissivity that is more than two orders of
magnitude greater than that of radio halos. On the other hand, the
mini-halos in cooling core clusters like A2029, Ophiuchus, and A1835
(Govoni et al. 2009) have a radio emissivity that is much more
typical of halos in merging clusters rather than to the mini-halos
previously known.  A1689 populates the same region of the $I_{\rm
  0}-r_{\rm e}$ plane of the other radio halos known in the
literature.  Although the e-folding radius of A1689 is not
particularly extended, it is larger than the e-folding radius of the
other mini-halos (they typically have $r_{\rm e}$$\lesssim 100$\,{\rm
  kpc}).  Therefore, the physical properties of the
diffuse emission in A1689 seem in good agreement with the
extrapolation of the properties of the other radio halos known in the
literature.

\subsection{Comparison of thermal and nonthermal emission}
\begin{figure*}
\centering
\includegraphics[width=18cm]{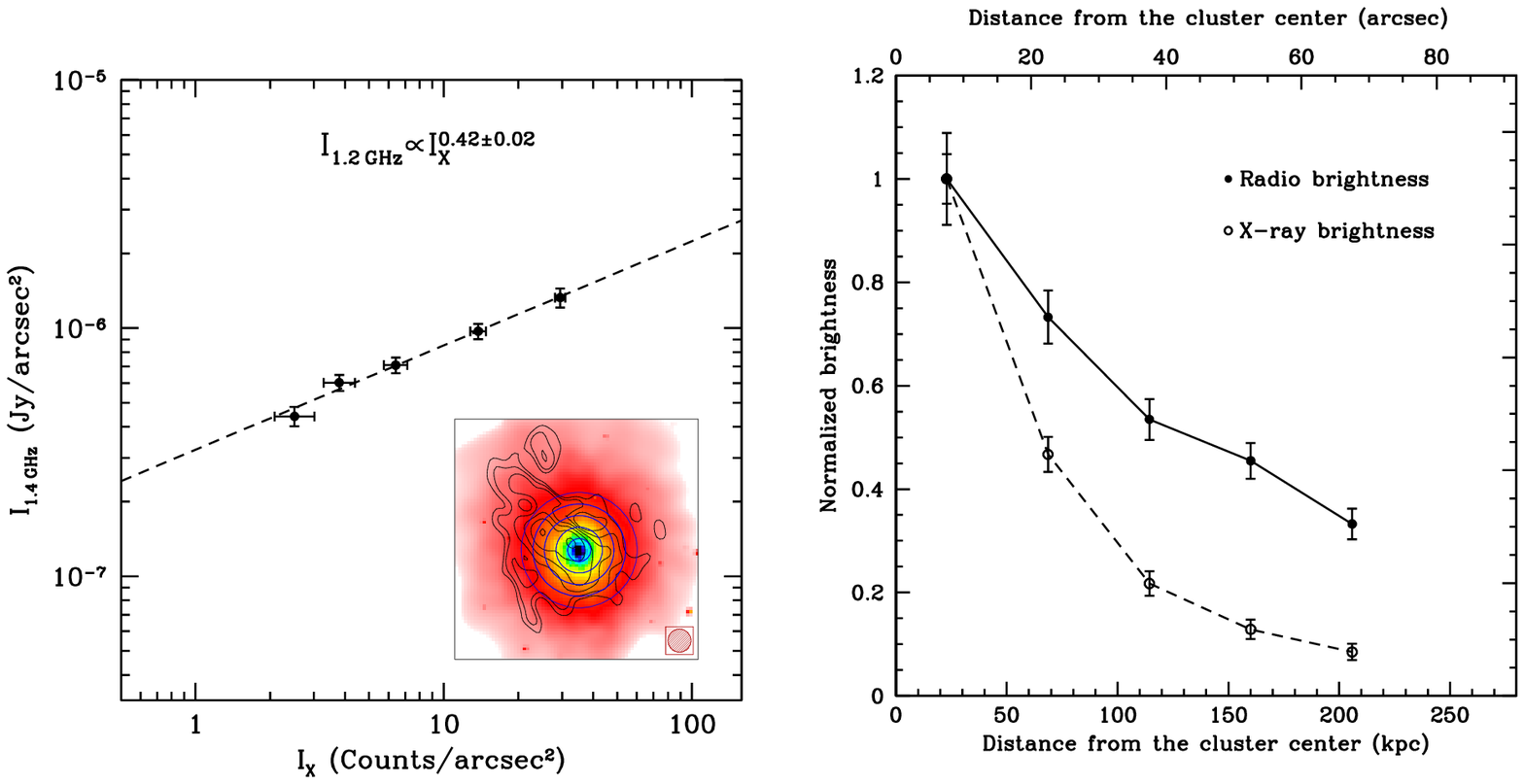}
\caption{\emph{Left}: Azimuthal average of the radio brightness at 1.2\,GHz versus the \emph{Chandra} X-ray image in A1689. Each dot represents the average of the emission in an annulus of half beam-width size (15\arcsec) centered on the cluster X-ray peak, as shown in the inset.
\emph{Right}: Normalized radio and X-ray radial profiles.
}
\label{radiovsx}
\end{figure*}
To investigate the possible presence of a radio - X-ray correlation in A1689,
in Fig.\,\ref{radiovsx} we present a quantitative comparison between the radio 
halo emission at 1.2\,GHz after the correction for the primary beam 
attenuation and the \emph{Chandra} X-ray image in the 0.1$-$10\,{\rm keV} 
band. Only regions where the radio halo emission is higher 
than a level of 3$\sigma$ have been considered. 
The radial profile was produced by using the same concentric annuli presented in Fig.\,\ref{profilo}. 
Each dot represents the average emission in each annulus. 
By fitting the trend with a power law relation
\begin{equation}
 I_{\rm 1.2\,GHz}\propto I_{\rm X}^{\rm a},
\end{equation} 
we obtained ${\rm a}$=0.42$\pm$0.02 (Fig.\,\ref{radiovsx}, \emph{left panel}), 
reinforcing the turbulence model of particle acceleration (e.g., Xu et al. 2010). 
The sublinear slope comes from the radial profile of the radio emission
being flatter than the X-ray one, as shown in the \emph{righthand panel} of Fig.\,\ref{radiovsx}, 
as in Coma and in A2319 (Govoni et al. 2001, Brown \& Rudnick 2011). 
A similar analysis was performed for the mini-halo in the 
Ophiucus cluster, where a superlinear correlation has instead been observed 
(Govoni et al. 2009).

\section{Discussion and conclusion}
\label{Conclusion}

As a part of an ongoing program aimed at finding new diffuse sources
in galaxy clusters, we have investigated the radio emission of the complex
galaxy cluster A1689.  We analyzed deep radio observations
carried out with the Very Large Array at 1.2 and 1.4\,{\rm GHz} in different
configurations.  To properly map these diffuse sources, high
sensitivity to the extended features is needed, along with modest
($\sim$20\arcsec) angular resolution in order to distinguish a real
diffuse source from the blend of unrelated discrete sources.  The
availability of sensitive radio images at different resolutions
revealed that at the center of A1689 a few discrete sources are
surrounded by a diffuse low surface brightness emission associated
with the intracluster medium.

We studied the morphological and physical radio properties
(i.e., length scale, central brightness, average emissivity) of the
diffuse emission in A1689 in comparison with that of other halos known
in the literature, by fitting their azimuthally averaged brightness
profile with an exponential law.  The best-fit of the exponential
model at 1.2\,{\rm GHz} yields a central brightness of 
$I_{\rm 0}$=1.7\,$\mu${\rm  Jy/arcsec}$^{\rm 2}$ and $r_{\rm e}=149$\,{\rm kpc}.  
From the central brightness and the e-folding radius, we derived an average
radio emissivity of $\langle J \rangle=1.7\times10^{\rm -42}$\,{\rm
  ergs}$^{\rm-1}${\rm cm}$^{\rm -3}${\rm Hz}$^{\rm -1}$, consistent
with that of all the other radio halo clusters known to date.
Moreover, we investigated the possible presence of a 
radio - X-ray correlation in A1689. 
We find a sublinear slope, which comes from the radial profile of the radio 
emission being flatter than the X-ray one.

The presence of a radio halo in A1689 strongly supports the 
cluster merger scenario discussed by Leonard et al. (2011). Detecting 
a diffuse nonthermal emission in a high-mass, merging structure  
agrees with the
evidence that radio halos are mostly found in massive merging systems.

\begin{acknowledgements}
We thank the referee for the very useful comments that helped to 
improve this paper. This research was partially supported by PRIN-INAF2009.
We acknowledge financial contribution from the agreement ASI-INAF I/009/10/0.
We thank Annalisa Bonafede and Roberto Pizzo for useful 
discussions.
V.V. acknowledges the hospitality of the Department of
Physics and Astronomy at the University of New Mexico, where part of this 
work was done.
GBT acknowledges support provided by the National Aeronautics and Space
Administration 
through Chandra Award Numbers GO0-11139X and GO0-11138B issued by the Chandra 
X-ray Observatory Center, which is operated by the Smithsonian Astrophysical
Observatory for and on behalf of the National Aeronautics Space Administration 
under contract NAS8-03060.
The National
Radio Astronomy Observatory (NRAO) is a facility of the National
Science Foundation, operated under cooperative agreement by Associated
Universities, Inc.  Funding for the SDSS and SDSS-II has been provided
by the Alfred P. Sloan Foundation, the Participating Institutions, the
National Science Foundation, the U.S. Department of Energy, the
National Aeronautics and Space Administration, the Japanese
Monbukagakusho, the Max Planck Society, and the Higher Education
Funding Council for England. The SDSS Web Site is
http://www.sdss.org/.  This research made use of Montage, funded by
the National Aeronautics and Space Administration's Earth Science
Technology Office, Computational Technologies Project, under
Cooperative Agreement Number NCC5-626 between NASA and the California
Institute of Technology. The code is maintained by the NASA/IPAC
Infrared Science Archive.
\end{acknowledgements}


\begin{thebibliography}{}

\bibitem{} Andersson, K.~E., \& Madejski, G.~M.\ 2004, \apj, 607, 190 
\bibitem{} Bacchi M., Feretti L., Giovannini G., Govoni, F.\ 2003, A\&A, 400, 465 
\bibitem{} Bonafede, A., et al. 2009, A\&A, 503, 707
\bibitem{} Broadhurst, T., Takada, M., Umetsu, K., Kong, X., Arimoto, N., Chiba, M., \& Futamase, T.\ 2005a, \apjl, 619, L143 
\bibitem{} Broadhurst, T., et al.\ 2005b, \apj, 621, 53 
\bibitem{} Brown, S., Duesterhoeft, J., \& Rudnick, L.\ 2011, \apjl, 727, L25
\bibitem{} Brown, S., \& Rudnick, L.\ 2011, \mnras, 412, 2
\bibitem{} Brunetti, G., \& Lazarian, A.\ 2007, MNRAS, 378, 245
\bibitem{} Brunetti, G., Venturi, T., Dallacasa, D., Cassano, R., Dolag, K., Giacintucci, S., \& Setti, G.\ 2007, \apjl, 670, L5 
\bibitem{} Buote, D.~A.\ 2001, ApJ, 553, L15 
\bibitem{} Burns, J.O., Sulkanen, M.E., Gisler, G.R., \& Perley, R.A.\ 1992, \apjl, 388, L49 
\bibitem{} Burns, J.~O., Hallman, E.~J., Gantner, B., et al.\ 2008, \apj, 675, 1125 
\bibitem{} Cassano, R., Ettori, S., Giacintucci, S., Brunetti, G., Markevitch, M., Venturi, T., \& Gitti, M.\ 2010, \apjl, 721, L82 
\bibitem{} Chen, Y., Reiprich, T.~H., B{\"o}hringer, H., Ikebe, Y., \& Zhang, Y.-Y.\ 2007, \aap, 466, 805 
\bibitem{} Clowe, D., \& Schneider, P.\ 2001, \aap, 379, 384 
\bibitem{} Coe, D., Ben{\'{\i}}tez, N., Broadhurst, T., \& Moustakas, L.~A.\ 2010, \apj, 723, 1678 
\bibitem{} Colafrancesco, S., 1999, Diffuse Thermal and Relativistic Plasma in Galaxy Clusters, 269
\bibitem{} Ensslin, T.~A., Biermann, P.~L., Klein, U., \& Kohle, S.\ 1998, \aap, 332, 395 
\bibitem{} Feretti, L.\ 2002, The Universe at Low Radio Frequencies, Proceedings of IAU Symposium 199, Pune, India. Edited by A. Pramesh Rao, G. Swarup, and Gopal-Krishna, 199, 133
\bibitem{} Feretti, L., Giovannini, G. 2008, A Pan-Chr. View of Clust. of Galax. and the Large-Scale Struct., 740, 143
\bibitem{} Ferrari, C., Govoni, F., Schindler, et al. 2008, Space Science Reviews, 134, 93
\bibitem{} Giovannini, G., \& Feretti, L.\ 2000, New Astronomy, 5, 335 
\bibitem{} Giovannini, G., Bonafede, A., Feretti, L., et al. \ 2009, A\&A, 507, 1257
\bibitem{} Giovannini, G., Feretti, L., Girardi, M., Govoni, F., Murgia, M., Vacca, V., \& Bagchi, J.\ 2011, \aap, 530, L5 
\bibitem{} Girardi, M., Fadda, D., Escalera, E., Giuricin, G., Mardirossian, F., \& Mezzetti, M.\ 1997, \apj, 490, 56 
\bibitem{} Gitti, M., Brunetti, G., Feretti, L., \& Setti, G.\ 2004, \aap, 417, 1 
\bibitem{} Gitti, M., Ferrari, C., Domainko, W.,  et al.\ 2007, \aap, 470, L25
\bibitem{} Govoni, F., En{\ss}lin, T.~A., Feretti, L., \& Giovannini, G.\ 2001, \aap, 369, 441 
\bibitem{} Govoni, F., Markevitch, M., Vikhlinin, A., et al.\ 2004, ApJ, 605, 695
\bibitem{} Govoni, F., Murgia M., Markevitch, M.,  et al.\ 2009, \aap, 499, 371 
\bibitem{} Govoni, F., Murgia, M., Giovannini, G., Vacca, V., \& Bonafede, A.\ 2011, \aap, 529, A69 
\bibitem{} Kawahara, H., Suto, Y., Kitayama, T., Sasaki, S., Shimizu, M., Rasia, E., \& Dolag, K.\ 2007, \apj, 659, 257 
\bibitem{} Lemze, D., Barkana, R., Broadhurst, T.~J., \& Rephaeli, Y.\ 2008, \mnras, 386, 1092 
\bibitem{} Leonard, A., King, L.~J., \& Goldberg, D.~M.\ 2011, \mnras, 413, 789 
\bibitem{} Liang, H., 1999, Diffuse Thermal and Relativistic Plasma in Galaxy Clusters, 33
\bibitem{} Limousin, M., et al.\ 2007, \apj, 668, 643 
\bibitem{} Miralda-Escude, J., \& Babul, A.\ 1995, \apj, 449, 18 
\bibitem{} Murgia, M., Govoni, F., Feretti, L., et al.\ 2004, \aap, 424, 429 
\bibitem{} Murgia, M., Govoni, F., Markevitch, M., et al.\ 2009, \aap, 499, 679 
\bibitem{} Murgia, M., Govoni, F., Feretti, L., \& Giovannini, G.\ 2010, \aap, 509, A86 
\bibitem{} Pedlar, A., Ghataure, H.~S., Davies, R.~D., et al.\ 1990, \mnras, 246, 477 
\bibitem{} Peng, E.-H., Andersson, K., Bautz, M.~W., \& Garmire, G.~P.\ 2009, \apj, 701, 1283 
\bibitem{} Peres, C.~B., Fabian, A.~C., Edge, A.~C., Allen, S.~W., Johnstone, R.~M., \& White, D.~A.\ 1998, \mnras, 298, 416 
\bibitem{} Reiprich, T.~H., B{\"o}hringer, H.\ 2002, \apj, 567, 716 
\bibitem{} Riemer-S{\o}rensen, S., Paraficz, D., Ferreira, D.~D.~M., Pedersen, K., Limousin, M., \& Dahle, H.\ 2009, \apj, 693, 1570 
\bibitem{} Rossetti, M., Eckert, D., Cavalleri, B.~M., et al.\ 2011, \aap, 532, A123
\bibitem{} Roettiger, K., Stone, J.~M., \& Burns, J.~O.\ 1999, ApJ, 518, 594
\bibitem{} Sarazin, C.~L.\ 2002, Merging Processes in Galaxy Clusters, Edited by L. Feretti, I.M. Gioia, G. Giovannini. Astrophysics and Space Science Library, Kluwer Academic Publishers, Dordrecht, 272, 1
\bibitem{} Skillman, S.~W., O'Shea, B.~W., Hallman, E.~J., Burns, J.~O., 
\& Norman, M.~L.\ 2008, \apj, 689, 1063 
\bibitem{} Skillman, S.~W., Hallman, E.~J., O'Shea, B.~W., Burns, J.~O., Smith, B.~D., \& Turk, M.~J.\ 2011, \apj, 735, 96 
\bibitem{} Struble, M.~F., \& Rood, H.~J.\ 1999, \apjs, 125, 35 
\bibitem{} Tyson, J.~A., Wenk, R.~A., \& Valdes, F.\ 1990, \apjl, 349, L1 
\bibitem{} Umetsu, K., \& Broadhurst, T.\ 2008, \apj, 684, 177 
\bibitem{} White, D.~A., \& Fabian, A.~C.\ 1995, \mnras, 273, 72 
\bibitem{} Xu, H., Li, H., Collins, D.~C., Li, S., \& Norman, M.~L.\ 2010, \apj, 725, 2152 






\end{thebibliography}
\end{document}